\documentclass[prl,twocolumn,superscriptaddress]{revtex4-2}
\usepackage{graphicx}
\usepackage{amssymb}
\usepackage{amsfonts}
\usepackage{amsmath}
\usepackage{lmodern}
\usepackage{mathrsfs}
\usepackage{mathdots}
\usepackage{collref}
\usepackage[colorlinks=true, citecolor=blue, urlcolor=blue]{hyperref}
\usepackage{tikz}
\usepackage{bm}
\renewcommand{\section}[1]{{\par\it #1.---}\ignorespaces}

\begin{document}

\title{Temperature uncertainty relation in non-equilibrium thermodynamics}
\author{Ning Zhang}
\affiliation{Lanzhou Center for Theoretical Physics and Key Laboratory of Theoretical Physics of Gansu Province, Lanzhou University, Lanzhou 730000, China}

\author{Si-Yuan Bai}
\affiliation{Lanzhou Center for Theoretical Physics and Key Laboratory of Theoretical Physics of Gansu Province, Lanzhou University, Lanzhou 730000, China}

\author{Chong Chen}\email{chongchenn@gmail.com}
\affiliation{Department of Physics and The Hong Kong Institute of Quantum Information of Science and Technology, The Chinese University of Hong Kong, Shatin, New Territories, Hong Kong, China}

\begin{abstract}
	Temperature uncertainty of a quantum system in canonical ensemble is inversely determined by its energy fluctuation, which is known as the temperature-energy uncertainty relation.  No such uncertainty relation was discovered for a non-equilibrium open quantum system. In this article, we derive a universal temperature uncertainty relation for general non-equilibrium processes. We find that it is the fluctuation of heat, which is defined as the change in bath energy, determines the temperature uncertainty in non-equilibrium case. Specifically, the heat is divided into  trajectory heat and  backaction heat, which are associated with the system's trajectory of evolution and the backaction of partial measurement on system, respectively. Based on this decomposition, we reveal that both correlations between system and bath's process function and state function are the resources for enhancing temperature precision. Our findings are conductive to design ultrahigh sensitive quantum thermometer.
\end{abstract}
\maketitle
\section{Introduction}
With decreasing system's size to nanoscale, values of physical quantities become fluctuated \cite{Horodecki2013,Rossnagel2016}. According to quantum mechanics, the fundamental accuracy of one physical quantity is always inversely determined by the fluctuation of another complementary quantity, which is known as the uncertainty principle \cite{Robertson1929,Friedland2013}. One famous example is the position and momentum uncertainty relation discovered by Heisenberg \cite{Heisenberg1927}. The information theory extends uncertainty relation to general cases, e.g., time and energy uncertainty relation \cite{Caneva2009,Taddei2013,Pires2016},  entropic uncertainty relations \cite{Deutsch1983,Maassen1988,Coles2017}, thermodynamic uncertainty relations \cite{Barato2015, Gingrich2016, Horowitz2020}, and temperature-energy uncertainty relation \cite{Velazquez2009}.

By taking the temperature as a state variable imprinted by thermal bath, temperature uncertainty relation focuses on the precision in temperature estimate of a small quantum system \cite{Gilmore1985,DePasquale2016,Miller2018}. The system can be in equilibrium or not.  In addition to a fundamental physical concept, temperature uncertainty relation is a resource theory for quantum thermometry \cite{Giazotto2006,Stace2010,Hofer2017,DePasquale2018}. With the developing of nanotechnologies, the temperature measuring on nanoscale and micro-kelvin precision become relevant \cite{Weld2009,Neumann2013,Kucsko2013,FerreiroVila2021}. The study of temperature uncertainty relation would help us figure out the resources for enhancing temperature precision and then design ultrahigh sensitive thermometer \cite{Hovhannisyan2021}. As the temperature sensing are based on both equilibrium states \cite{Correa2015,DePasquale2016}  and non-equilibrium states \cite{Seah2019,Mukherjee2019,Mitchison2020}, the study of temperature uncertainty relation for non-equilibrium processes become relevant.

Consider a quantum system described by the temperature dependent state $\rho_{\beta}$ with $\beta=(k_B T)^{-1}$ being the inverse temperature. The probability distribution function of outputs by measuring observable $\hat{x}$ is denoted by $p(x;\beta')$, where $x$ is the measured value of $\hat{x}$ and $\beta^{\prime}$ is the estimated temperature. Generally, $\beta'$ not equals to $\beta$ due to the fluctuation. The distinguishability between $p(x;\beta')$ and the true one $p(x;\beta)$ is described by the relative entropy \cite{Kullback1951}
\begin{equation}
	S(p(x; \beta')||p(x; \beta))=\int dx  p(x;\beta') \ln\frac{p(x;\beta')}{p(x;\beta)}.
	\label{eq:RE}
\end{equation}
The probability functional of all the possible probability functions $p(x,\beta')$ reads $\mathbb{P}_{p(x;\beta')} \propto e^{-\nu S(p(x;\beta')||p(x;\beta))}$, where $\nu$ is the number of repeat measurements \cite{Vedral2002}. A possible expansion of $\beta'$ around $\beta$ shows that
\begin{equation}
	S(p(x;\beta')||p(x;\beta))= (\beta'-\beta)^{2} F_{\beta} +\mathcal{O}(\beta'-\beta)^{3},
	\label{eq:FisherInformation}
\end{equation}
where $F_{\beta}= \int dx p(x;\beta)L_\beta^2(x)$ is the  Fisher information with $L_\beta(x)\equiv \partial_{-\beta} \ln p(x;\beta)$ being the score \cite{Fisher1925}. Then up to the second order, the Gaussian form of $\mathbb{P}_{p(x;\beta')}$ indicates that uncertainty of temperature $\beta$ is inversely determined by the Fisher information or the fluctuation of the score,  such that
\begin{equation}
	\Delta \beta^{2} \Delta L_\beta^{2} \ge 1,
\label{eq:UncertaintyR}
\end{equation}
where we set $\nu=1$.  This is the general temperature uncertainty inequality or known as the Cram\'{e}r-Rao bound in metrology theory \cite{Braunstein1996}. The above analyses can be easily generalized to the case that without any prefer measuring operator $\hat{x}$ \cite{Braunstein1994}, where the score is replaced by the logarithmic derivative operator $\hat{L}_\beta$ with definition
\begin{equation}
	\frac{d\rho_{\beta}}{d-\beta}=\{\hat{L}_\beta,\rho_{\beta}\}.
	\label{eq:LDO}
\end{equation}
$\{\hat{A},\hat{B}\}=\frac{1}{2}(\hat{A}\hat{B}+\hat{B}\hat{A})$ is the anti-commutation operation and $(-\beta)$ is taken for convenience. The fluctuation $\Delta L_\beta^{2}$ is then recast to the quantum Fisher information (QFI)
\begin{equation}
\Delta L^{2}_{\beta}=\text{Tr}[\hat{L}_\beta^{2} \rho_{\beta}].
\label{eq:QFI-L}
\end{equation}

Although Eq. \eqref{eq:UncertaintyR} and \eqref{eq:QFI-L} build a general uncertainty inequality between the temperature uncertainty and  the fluctuation of $\hat{L}_\beta$, the absence of physical interpretation of $\hat{L}_\beta$ prevents it to be a uncertainty relation like position-momentum uncertainty or time-energy uncertainty. Associating the information operator $\hat{L}_{\beta}$ with a physical quantity is a challenging and important task, which not only reveals the physical resources limiting the temperature uncertainty,  but also builds a relationship between information theory and quantum thermodynamics \cite{Jaynes1957}. When the quantum system is weakly coupled with a thermal bath, the steady state is described by the canonical ensemble, i. e., $\rho_{\infty}=e^{-\beta\hat{H}_S}/\text{Tr}[e^{-\beta\hat{H}_S}]$. One can readily find that $\hat{L}_\beta=\hat{H}_S-\langle \hat{H}_S\rangle$. Then Eq. \eqref{eq:UncertaintyR} is recast to the well-known temperature-energy uncertainty relation \cite{Landau1980, Paris2015}. Beyond the weak system-bath coupling, the equilibrium case was studied in Ref. \cite{Miller2018}. For non-equilibrium cases,  to the best of our knowledge, no such relation was discovered even in weak coupling limit. Here, we aim to establish a general temperature uncertainty relation in non-equilibrium thermodynamics.

\section{Phase-space formulation}
To obtain a general temperature uncertainty relation, we resort to the phase-space formulation of quantum mechanics. The quasiprobability distribution for a state $\rho$ is defined through the generalized Weyl rule as \cite{Brif1999, Rundle2021}
\begin{equation}
	C^{(s)}_{\rho}(\Omega)\equiv\text{Tr}[\rho \hat{\Pi}^{(s)}_{\Omega}],~(s=\pm 1),
	\label{eq:QuasiProb}
\end{equation}
where $\hat{\Pi}^{(-1)}_{\Omega}=|\Omega\rangle\langle \Omega|$ satisfies $\int d\mu(\Omega)\hat{\Pi}^{(-1)}_{\Omega}=1$ with $|\Omega\rangle$ being the generalized coherent state of the system and $d\mu(\Omega)$ being the integration measure.  The expression of $\hat{\Pi}^{(1)}_{\Omega}$ depends on the specific system. For harmonic oscillator, one has $\hat{\Pi}^{(1)}_{\alpha}=\int \frac{d^2 \xi}{\pi} e^{\xi (\hat{a}^{\dagger}-\alpha^*)-h.c.}$ in coherent state basis, where $\alpha$ is a complex number and $\hat{a}$ is the annihilation operator \cite{Carmichael1999}. Its form for a spin is given in Ref. \cite{Brif1999}. The inverse Weyl rule reads
\begin{equation}
	\rho = \int d\mu(\Omega) {C}^{(s)}_{\rho}(\Omega) \hat{\Pi}^{(-s)}_{\Omega}.
	\label{eq:QuasiRep}
\end{equation}
Note that similar representation for any operator $\hat{O}$  is defined as $C^{(s)}_{O}(\Omega)=\text{Tr}[\hat{O} \hat{\Pi}^{(s)}_{\Omega}]$.
It can be found that $C^{(1)}_{\rho}(\Omega)$ and $C^{(-1)}_{\rho}(\Omega)$ are the $P$- and $Q$-functions of $\rho$, respectively. In the following, we will denote them as $P(\Omega)$ and $Q(\Omega)$.

Our first step is to find an expression of $\Delta L^2_{\beta}$ in term of the score of $Q(\Omega)$, which is defined as $\mathcal{L}_{\beta}(\Omega)=\partial_{-\beta}\ln Q(\Omega)$. Combining the formula $\Delta L_\beta^{2}=\text{Tr}[ (\partial_{-\beta} \rho) \hat{L}_\beta]$ with the $P$-representation of $\hat{L}_\beta$, noted as $C^{(1)}_{L_\beta}(\Omega)$, we get $\Delta L_\beta^{2}=\int d\mu(\Omega)\mathcal{L}_\beta(\Omega)Q(\Omega) C^{(1)}_{L_\beta}(\Omega)$. A transformation between $\mathcal{L}_{\beta}(\Omega)$ and $C^{(1)}_{L_\beta}(\Omega)$ is derived from Eq. \eqref{eq:LDO} that $\mathcal{L}_\beta(\Omega)Q(\Omega)=\int d\mu(\Omega')\mathcal{T}(\Omega,\Omega'){C}^{(1)}_{L_\beta}(\Omega^\prime)$,
where $\mathcal{T}(\Omega,\Omega^{\prime})=\text{Tr}[\{\hat{\Pi}^{(-1)}_{\Omega}, \hat{\Pi}^{(-1)}_{\Omega^{\prime}}\}\rho]$  (see Ref. \cite{SupplementalMaterial}). By introducing  the inverse transformation $\mathcal{T}^{-1}$ satisfying $\int d\mu(\Omega^{''}) \mathcal{T}^{-1}(\Omega,\Omega^{''})\mathcal{T}(\Omega^{''},\Omega^{'})=\delta(\Omega-\Omega^{\prime})$, we finally obtain
\begin{equation}
	\Delta L_\beta^{2}= \int d \mu(\Omega) d\mu(\Omega') g^{\Omega, \Omega^{\prime}} \mathcal{L}_\beta(\Omega) \mathcal{L}_\beta(\Omega'),
	\label{eq:QFI-PS}
\end{equation}
where $g^{\Omega, \Omega'}\equiv Q(\Omega)\mathcal{T}^{-1}(\Omega,\Omega')Q(\Omega^{\prime})$ is symmetric and positive-defined (see Ref. \cite{SupplementalMaterial}).

\begin{figure}[thpb]
  \centering
  \includegraphics[width=1.0\columnwidth]{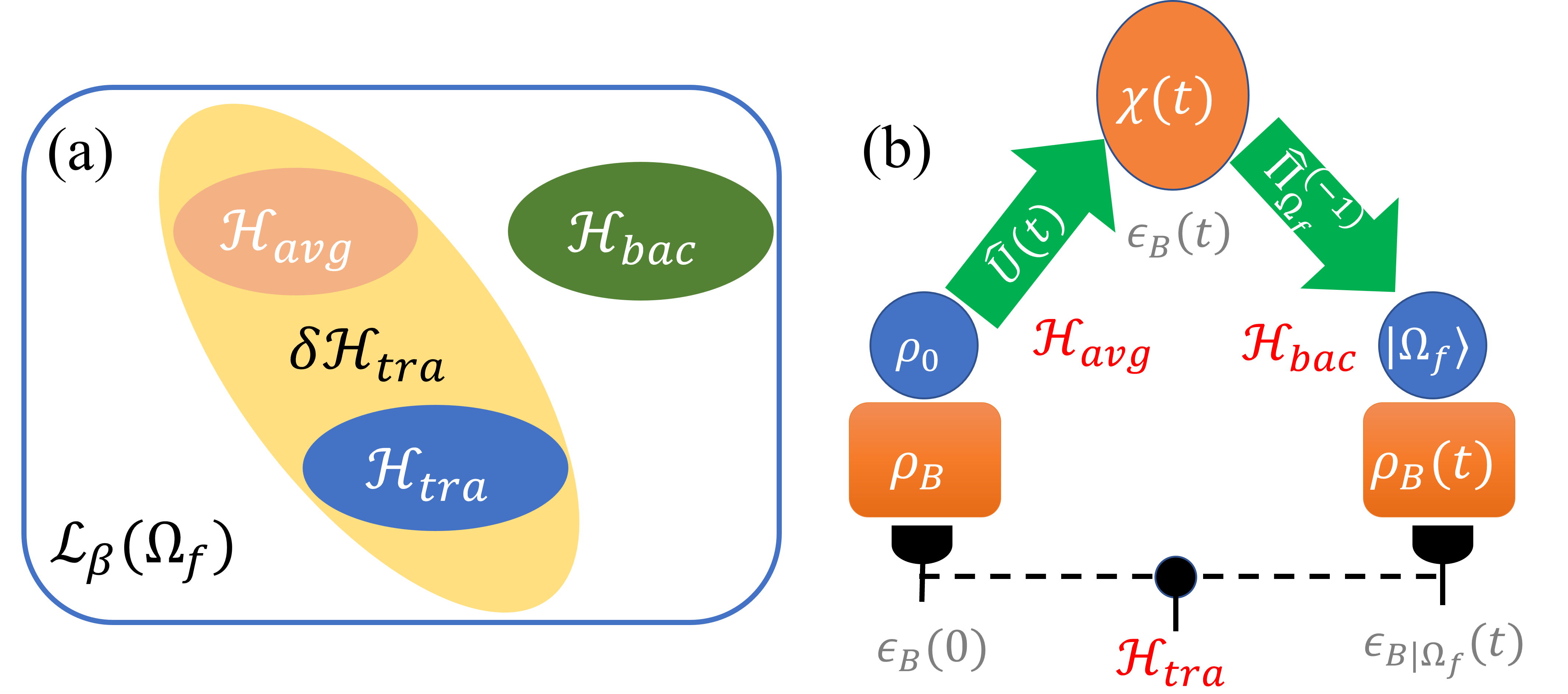}
  \caption{(a) Decomposition of the score as $\mathcal{L}_\beta(\Omega_f)=\delta \mathcal{H}_{tra}+\mathcal{H}_{bac}$ with $\delta \mathcal{H}_{tra}=\mathcal{H}_{tra}-\mathcal{H}_{avg}$; (b) The initial product state $\chi(0)=\rho_0 \otimes \rho_{B}$ evolves to $\chi(t)$ under unitary evolution $\hat{U}_{t}$. A following projective measurement on system reduces the total state to the final state $|\Omega_f \rangle \langle \Omega_f|\otimes \rho_B(t)$ with probability $Q_t(\Omega_f)$.   Sequential measurements on bath energy of initial state and final state yields $\epsilon_{B}(0)$ and $\epsilon_{B|\Omega_f}(t)$, respectively. The average of $\epsilon_{B|\Omega_f}(t)$ over different $\Omega_f$ gives $\epsilon_B(t)$.  Differences of mean values $ \overline{\epsilon_B(0)}- \overline{ \epsilon_B(t)}$ and $ \overline{\epsilon_{B|\Omega_f}(t)}- \overline{ \epsilon_{B}(t)}$ are defined as average heat $\mathcal{H}_{avg}$ and backation heat $\mathcal{H}_{bac}$, respectively. At last, the joint detection of initial and final states yields the trajectory heat $\mathcal{H}_{tra}\equiv\overline{\epsilon_{B|\Omega_f}(t)-\epsilon_{B}(0)}$. }
  \label{fig:Decomposition}
\end{figure}

As our first main result, Eq. \eqref{eq:QFI-PS} provides us an exact phase-space expression of QFI. By noting $g^{\Omega, \Omega'}$ is symmetric and positive defined, we take it as a metric in the phase space. Then QFI is recast to the variance of $\mathcal{L}_\beta(\Omega)$ under metric $g$. In contrast to the quantum operator $\hat{L}_{\beta}$, $\mathcal{L}_{\beta}$ is a classical quantity without any quantum property. Here all quantum effects in QFI are incorporated in metric $g$. In the classical limit $\hbar \rightarrow 0$, one has $g^{\Omega,\Omega^{\prime}} =\delta (\Omega-\Omega') Q(\Omega)$ and hence $\Delta L^{2}_{\beta} = \int d\mu(\Omega) \mathcal{L}^{2}_\beta(\Omega)Q(\Omega)$ recovers the classical Fisher information.  Through converting the elusive QFI to the variance of the classical quantity $\mathcal{L}_\beta$,  Eq. \eqref{eq:QFI-PS} makes the physical interpretation of $\Delta L^{2}_{\beta}$ possible.  In the following, we will apply this result to a general non-equilibrium dynamics.

\section{General temperature uncertainty relation}
Consider a general open quantum system (set $\hbar=1$)
\begin{equation}
	\hat{H}=\hat{H}_{S}+\hat{H}_{B}+\hat{H}_{I},
	\label{eq:Hamil}
\end{equation}
where $\hat{H}_{S}$ is the system's Hamiltonian, $\hat{H}_B$ is the bath Hamiltonian, and $\hat{H}_{I}$ is system-bath interaction. The system and bath are decoupled at time $t\le0$ with an initial product state $\chi(0)=\rho_{0}\otimes\rho_B$. Here $\rho_B=e^{-\beta \hat{H}_B}/Z_B$ is a thermal state with $Z_B=\text{Tr}[e^{-\beta \hat{H}_B}]$. It is the unitary evolution $\chi(t)=\hat{U}_{t}\chi(0) \hat{U}^{\dagger}_{t}$ with $\hat{U}_t=e^{-i \hat{H }t}$ that encodes bath's temperature information into the system's state $\rho_t=\text{Tr}_{B}[\chi(t)]$. The corresponding $Q$-representation of $\rho_t$ reads $Q_{t}(\Omega_{f})=\text{Tr}[\hat{\Pi}^{(-1)}_{\Omega_f} \chi(t)]$. From it, we derive the score as
\begin{equation}
\mathcal{L}_\beta(\Omega_f)=\frac{1}{Q_{t}(\Omega_{f})} \text{Tr}[\hat{\Pi}^{(-1)}_{\Omega_f} \hat{U}_t \hat{H}_B \chi(0)\hat{U}_t^\dag]-\text{Tr} [\hat{H}_B \chi(0)].
\label{eq:heat}
\end{equation}

A clever decomposition shows that $\mathcal{L}_\beta(\Omega_f)$ is the sum of trajectory heat, backaction heat, and average heat, i. e., $\mathcal{L}_\beta(\Omega_f) = \mathcal{H}_{tra} + \mathcal{H}_{bac}-\mathcal{H}_{avg}$ with detail expressions
\begin{eqnarray}
\label{eq:decomposition-tra}
\mathcal{H}_{tra} &=& \frac{1}{Q_t(\Omega_f)} \text{Tr}[\hat{\Pi}^{(-1)}_{\Omega_f} (\hat{U}_t \hat{H}_B\chi(0)\hat{U}_t^\dag-\hat{H}_B \chi(t))],  ~~\\
\label{eq:decomposition-cor}
\mathcal{H}_{bac} &=&\frac{1}{Q_t(\Omega_f)}\text{Tr}[\hat{\Pi}^{(-1)}_{\Omega_f}  \hat{H}_B  \chi(t)]-\text{Tr}[ \hat{H}_B \chi(t)],  \\
\label{eq:decomposition-avg}
\mathcal{H}_{avg} &=&\text{Tr}[\hat{H}_B \chi(0)-\hat{H}_B \chi(t)].
\end{eqnarray}
Heat here is defined as the change of bath energy \cite{Goold2014,Popovic2021}. The physical meaning of these heat terms is explained via a schematic diagram shown in Fig. \ref{fig:Decomposition}. In such scheme, the total system evolves from a product state $\rho_0\otimes \rho_B$ to $\chi(t)$ under unitary evolution $\hat{U}(t)$, and finally reduces to $|\Omega_f\rangle \langle \Omega_f |\otimes \rho_B(t)$ with probability $Q_t(\Omega_f)$ under the projective measurement $\hat{\Pi}^{(-1)}_{\Omega_f}$ with $\Omega_f$ taking all the possible values in phase space \cite{Banaszek1996}. Twice sequential measurements of $\hat{H}_B$ are applied to the initial and final states, which yield the outputs $\epsilon_B(0)$ and $\epsilon_{B|\Omega_f}(t)$, respectively.  Above processes are repeated many times for output average. Bath energy $\epsilon_B(t)$ of the unprojected state $\chi(t)$ corresponds to the averaged $\epsilon_{B|\Omega_f}(t)$ over different $\Omega_f$. The relation between measurement outputs and different heat terms is shown below. The average of the difference between twice sequential measurements is defined as the trajectory heat, i. e., $\mathcal{H}_{tra}=\overline{\epsilon_{B}(0)- \epsilon_{B|\Omega_f}(t)}$. Here $\bar{\cdot}$ indicates the mean of repeated measurements. It follows the standard two-point measurements and reflects the loss of bath energy along system's given evolution \cite{Talkner2007, Aurell2018}. In experiments, it is measured by the joint detection shown in Fig. \ref{fig:Decomposition}(b). The difference between averaged $\epsilon_{B|\Omega}(t)$ and averaged $\epsilon_{B}(t)$ is defined as the backaction heat, i. e., $\mathcal{H}_{bac}=\overline{\epsilon_{B|\Omega_f}(t)}- \overline{\epsilon_{B}(t)}$. Being the difference of bath energy before and after the projective measurement $\hat{\Pi}^{(-1)}_{\Omega_f}$ on system, $\mathcal{H}_{bac}$ is the heat induced by the backaction of the projective measurement and reflects the correlation between system and bath \cite{Esposito2010,Planella2022}. It goes to zero when $\chi(t)$ is a product state, e.g., $\chi(t)=\rho_{t}\otimes \rho_B(t)$. At last, the bath energy difference before and after the unitary evolution is defined as the average heat $\mathcal{H}_{avg}= \overline{\epsilon_{B}(0)}-\overline{\epsilon_{B}(t)}$ \cite{Popovic2021}.

By further noting $\mathcal{H}_{avg}=\int d\mu(\Omega_f) \mathcal{H}_{tra}  Q_{t}(\Omega_f)$ is the mean of $\mathcal{H}_{tra}$, we get a simple expression of the score
 \begin{equation}
 \mathcal{L}_\beta(\Omega_f)= \delta \mathcal{H}_{tra}+\mathcal{H}_{bac},
 \label{eq:L-heat}
  \end{equation}
 where both $\delta \mathcal{H}_{tra}=\mathcal{H}_{tra} -\mathcal{H}_{avg}$ and $\mathcal{H}_{bac}$ are deviations because $\int d\mu(\Omega_f) \mathcal{H}_{bac}  Q_{t}(\Omega_f)=0 $.
From Eqs. (\ref{eq:decomposition-tra}-\ref{eq:decomposition-avg}), one can find that both $\delta \mathcal{H}_{tra}$ and $\mathcal{H}_{bac}$ are deviations caused by applying the projective measurement $\hat{\Pi}^{(-1)}_{\Omega_f}$ on system part, which reflect the influence of the system's status on heat and bath energy, respectively. Note that heat is a process function, rather than a state function like bath energy  \cite{Campisi2011}. Therefore, $\delta \mathcal{H}_{tar}$ and $\mathcal{H}_{bac}$ are related to correlations between system and bath's process function and state function, respectively.  A strong correlation between system and bath would cause large $\delta \mathcal{H}_{tra}$ and $\mathcal{H}_{bac}$ and then a large QFI.  Consequently the correlation between system and bath acts as the resources for decreasing the temperature uncertainty. Such a conclusion is agreement with strong coupling enhanced thermometry precision \cite{Correa2017, Hovhannisyan2018, Mehboudi2019, Planella2022}. Applying these results into Eq. \eqref{eq:QFI-PS}, we get
\begin{equation}
\Delta L_\beta^2= \langle (\delta \mathcal{H}_{tra}+\mathcal{H}_{bac})^2\rangle_g,
\label{eq:FOH}
\end{equation}
where the average is defined as $\langle O^2 \rangle_g \equiv \int d\mu(\Omega_f) d\mu(\Omega^\prime_f) g^{\Omega_f,\Omega^\prime_f} O(\Omega_f) O(\Omega^{\prime}_f)$.
 As the main result of this paper, Eq. \eqref{eq:FOH} reveals that it is the fluctuation of trajectory heat plus backaction heat under metric $g$ that fundamentally determines the temperature uncertainty, i. e.,
\begin{equation}
\Delta \beta^2 \langle (\delta \mathcal{H}_{tra}+\mathcal{H}_{bac})^2 \rangle_g \ge 1.
\label{eq:GTU}
\end{equation}
It is a general temperature uncertainty relation in non-equilibrium thermodynamics, applicable to any open quantum system.

\section{Example}
In the following, we use a quantum Brownian motion model to illustrate the temperature uncertainty relation. The Hamiltonian reads \cite{Caldeira1983}
\begin{equation}
\hat{H}=\omega_{0} \hat{a}^{\dagger} \hat{a}+\sum_k \omega_k \hat{b}^{\dagger}_{k} \hat{b}_{k} +(\hat{a}^{\dagger}+\hat{a})\sum_k g_k (\hat{b}^{\dagger}_{k} + \hat{b}_{k}),
\end{equation}
where $\hat{a}$ is the annihilation operator of system with  characteristic energy $\omega_0$, $g_{k}$ is the coupling strength, and $\hat{b}_{k}$ is the annihilation operator of $k$th bath mode with characteristic energy $\omega_{k}$.  The coherence state $|\alpha\rangle=\hat{D}(\alpha)|0\rangle$ is used for phase-space formulation with displacement operator $\hat{D}(\alpha)=e^{\alpha \hat{a}^\dagger-h.c.}$ and vacuum state $|0\rangle$. The corresponding integration measure reads $d\mu(\alpha)=\frac{d^2\alpha}{\pi}$.

Consider a coherent thermal initial state $\rho_0= \hat{D}(\alpha_0) e^{-\beta_0 \omega_0 \hat{a}^{\dagger}\hat{a}}\hat{D}^{\dagger}(\alpha_{0})/Z_0$, where $\beta_0$ is the initial inverse temperature and $Z_0=\frac{1}{1+\bar{n}_{0}}$ is normalization factor with $\bar{n}_0=\frac{1}{e^{\beta_0 \omega_0}-1}$. Under Markovian limit, $Q_{t}(\alpha_f)$ satisfies the Fokker-Planck equation \cite{Carmichael1999}
\begin{equation}
Q_{t}(\alpha_{f})=\frac{1}{1+\bar{n}(t)}e^{-\frac{|\alpha_{f}-\alpha_t|^{2}}{1+\bar{n}(t)}},
\end{equation}
where $\alpha_t=\alpha_0 e^{-i \omega_0 t-\gamma t/2}$ is the time-dependent displacement and $\bar{n}(t)=\bar{n}_0e^{-\gamma t}+\bar{n}_{\infty}(1-e^{-\gamma t})$ is the average excitation number with $\bar{n}_{\infty}=\frac{1}{e^{\beta \omega_0}-1}$ and $\gamma$ being decay rate. It describes the drift and diffusion processes of a Gaussian wave-package. The score is then derived as
\begin{equation}
\mathcal{L}_{\beta}= \omega_{0}  \frac{ \bar{n}_{\infty}(1+\bar{n}_{\infty})(1-e^{-\gamma t})}{(1+\bar{n}(t))^2} \delta |\alpha_{f}-\alpha_{t}|^{2},
\end{equation}
where $\delta |\alpha_f-\alpha_t|^2=|\alpha_f-\alpha_t|^2-(1+\bar{n}(t))$.

\begin{figure}
  \centering
  \includegraphics[width=1.0\columnwidth]{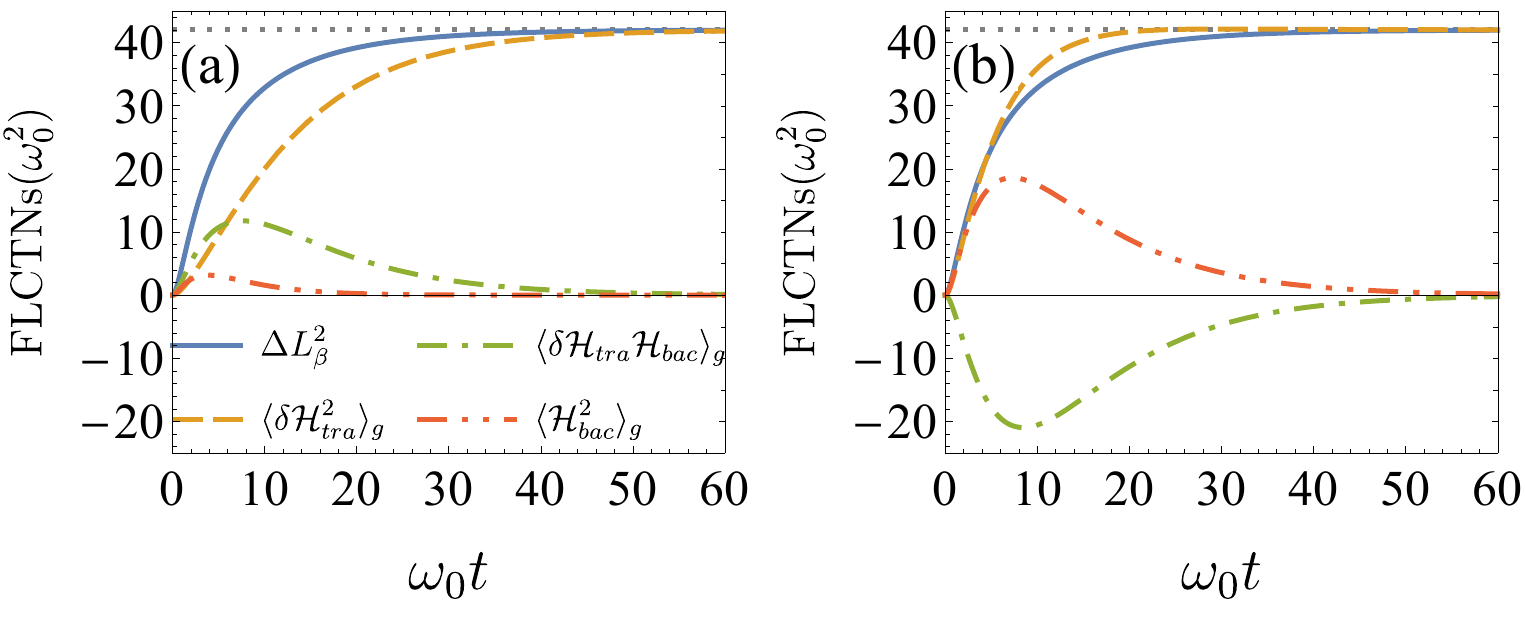}
  \caption{The evolution of  $\Delta L^{2}_\beta$ and the covariance matrix of trajectory heat deviation $\delta \mathcal{H}_{tra}$ and backaction heat $\mathcal{H}_{bac}$. (a) without initial coherence $\alpha_0=0$; (b) with initial coherence $\alpha_0=3.0$. Gray dotted line is  $\Delta H_S^2$ under equilibrium state.  Parameters are $\gamma=0.1\omega_0$, $\bar{n}_{0}=1.0$, and $\bar{n}_{\infty}=6.0$. Here we set $\omega_0=1.0$.   }
  \label{fig:fluctuation}
\end{figure}

An important property of Markovian process is that the interaction energy caused by $\hat{H}_{I}$ is ignorable, which makes the heat released by bath exactly equals to the energy absorbed by system \cite{Fei2018}. Based on this property, one can get the average heat $\mathcal{H}_{avg}=\omega_0 [\bar{n}(t)+|\alpha_t|^2-\bar{n}_{0}-|\alpha_0|^2]$ and the trajectory heat deviation (see Ref. \cite{SupplementalMaterial})
\begin{align}
\delta \mathcal{H}_{tra}=& \omega_{0}[(1-e^{-\gamma t}\frac{\bar{n}_{0}(1+\bar{n}_{0})}{\bar{n}(t)(1+\bar{n}(t))}) \frac{\bar{n}(t)}{1+\bar{n}(t)} \delta|\alpha_{f}-\alpha_t|^{2} \nonumber \\
&+ \frac{\bar{n}(t)-\bar{n}_{0}}{1+\bar{n}(t)} ((\alpha^{\ast}_{f}-\alpha^{\ast}_t)\alpha_{t}+h.c.)].
\end{align}
$\delta \mathcal{H}_{tra}$ reflects the correlation between system and bath's process function. At $t=0$, one has $\delta \mathcal{H}_{tra}=0$ since $\chi(0)$ is a product state. Although $\chi(\infty)=\rho_{\infty}\otimes \rho_B$ is also a product state, one has a nonzero deviation, $\delta \mathcal{H}_{tra}=\frac{\bar{n}_{\infty} \omega_0 }{1+\bar{n}_{\infty}}\delta|\alpha_{f}|^{2}$, which reflects the correlation between $\rho_{\infty}$ and bath's process function still exists. Here relations $\bar{n}(0)=\bar{n}_0$, $\bar{n}(\infty)=\bar{n}_{\infty}$, and $\alpha_\infty=0$ are used.

The corresponding backaction heat is obtained from the formula $\mathcal{H}_{bac}=\mathcal{L}_{\beta}(\alpha_f)-\delta\mathcal{H}_{tra}$, which reads
\begin{align}
\mathcal{H}_{bac}=&\omega_{0}\frac{\bar{n}(t)-\bar{n}_{0}}{1+\bar{n}(t)} [
\frac{\bar{n}_{\infty}-\bar{n}(t)}{1+\bar{n}(t)}\delta |\alpha_{f}-\alpha_{t}|^{2} \nonumber \\ &-((\alpha^{\ast}_{f}-\alpha^{\ast}_t)\alpha_{t}+h.c.)].
\end{align}
It reflects the correlation between system and bath's state function. The pre-factor $\frac{\bar{n}(t)-\bar{n}_0}{1+\bar{n}(t)}$ indicates that the correlation is built through the heat exchange. When system and bath are at the same temperature, one has $\bar{n}_{0}=\bar{n}_{\infty}=\bar{n}(t)$ and thus $\mathcal{H}_{bac}=0$. For the cases that  $\bar{n}_0 \neq \bar{n}_{\infty}$,  $\mathcal{H}_{bac}$ has a nonzero value except for $t=0$, or $\infty$, which indicates $\chi(t)$ is in general a correlated state.

The action of metric $g$ on $\mathcal{L}_{\beta}(\alpha_f)$ reads (see Ref. \cite{SupplementalMaterial})
\[\int \frac{d^2\alpha^{\prime}_f}{\pi} g^{\alpha_f, \alpha^{\prime}_f}\mathcal{L}_{\beta}(\alpha^{\prime}_f)=\frac{1+\bar{n}(t) }{\bar{n}(t)} \mathcal{L}_{\beta}(\alpha_f)Q_{t}(\alpha_f), \]
where the factor $\frac{1+\bar{n}(t) }{\bar{n}(t)}$ reflects the quantum modification. The fluctuation $\Delta L^2_{\beta}$ according to Eq. \eqref{eq:FOH} reads
\begin{equation}
  \Delta L^{2}_\beta=\omega^2_{0} \frac{\bar{n}^2_{\infty}(1+\bar{n}_{\infty})^2(1-e^{-\gamma t})^2}{\bar{n}(t)(1+\bar{n}(t))}.
\end{equation}
It is interesting to see that the result is $\alpha_0$ independent, which indicates the initial coherence has no contribution to temperature uncertainty.  The detail expression $\bar{n}(t)=\bar{n}_0 e^{-\gamma t}+\bar{n}_{\infty}(1-e^{-\gamma t})$ shows that $\Delta L^2_{\beta}$ decreases with increasing the system's initial temperature.  In the long time limit, we have $\Delta L^{2}_{\beta}=\omega^{2}_{0} \bar{n}_{\infty}(1+\bar{n}_{\infty})$, which recovers the energy fluctuation, i.e., $\Delta L^2_{\beta}= \Delta H^{2}_{S}$.

Evolutions of $\Delta L^2_{\beta}$ and covariance matrix of $\delta \mathcal{H}_{tra}$ and $\mathcal{H}_{bac}$ are shown in Fig. \ref{fig:fluctuation}. Both incoherent case with $\alpha_0= 0$ (shown in subfig. (a)) and coherent case (shown in subfig. (b)) are considered. Results show that $\Delta L^2_{\beta}$ is coherence independence, which monotonically increases with time evolution and relaxes to $\Delta H^2_{S}$.  In contrast, all covariances in covariance matrix show a coherence dependence. Specifically, coherence increase  $\langle \delta\mathcal{H}^2_{tra}\rangle_g$ and $\langle \mathcal{H}^2_{bac}\rangle_g$, whereas decreases the covariance $\langle \delta \mathcal{H}_{tra}\mathcal{H}_{bac}\rangle_g$, which makes $\Delta L^2_\beta$ is coherence independent. In addition, the variance of backaction heat $\langle \mathcal{H}^2_{bac} \rangle_g$ equals to zero at $t=0$ or $\infty$ and has a peaks at $t\approx 1/\gamma$, which indicates the system and bath are strongly correlated at $t\approx 1/\gamma$. In long time limit, only the element $\langle \delta \hat{H}^{2}_{tra}\rangle_g$ of covariance matrix has a finite value. Thus it is the correlation between system and bath's process function that determines the equilibrium state temperature uncertainty.

Beyond Markovian limit, the score $\mathcal{L}_\beta$, trajectory heat deviation $\delta \mathcal{H}_{tra}$, and backaction heat $\mathcal{H}_{bak}$ can be solved by using the path-integral influence functional method \cite{Funo2018, Funo2018a}. In contrast to the Markovian case, interaction energy is non-negligible, which makes equilibrium state $\langle \mathcal{H}^2_{bac}\rangle_{g} \neq 0 $ and $\Delta L^{2}_\beta$ different from the energy fluctuation $\Delta H^{2}_{S}$ \cite{Miller2018}.

\textit{Discussion and Conclusion--}
The non-Markovian equilibrium state temperature uncertainty relation was considered in Ref. \cite{Miller2018} by assuming $\rho_{\infty}= e^{-\beta \hat{H}^{\ast}_{S}}/Z^{\ast}_{S}$, where $\hat{H}^{\ast}_{S}$ is the system-bath interaction modified Hamiltonian \cite{Seifert2016}. Based on the Wigner-Yanase-Dyson skew information (SI), they find that the temperature uncertainty relation reads
\begin{equation}
  \Delta\beta^{2} (C_{S} -\langle \partial_{-\beta} \hat{E}^{\ast}_S \rangle) \ge 1,
  \label{eq:TU-SH}
\end{equation}
where $C_{S}\equiv \partial_{-\beta} \langle \hat{E}^{\ast}_{S} \rangle$ with energy operator $\hat{E}^{\ast}_{S} \equiv \partial_{\beta} (\beta \hat{H}^{\ast}_{S})$ is recognized as the heat capacity and $\langle \partial_{-\beta} \hat{E}^{\ast}_{S} \rangle$  is recognized as the extra dissipation term. Here we prove that this result can be recovered by our conclusion with using the phase-space formulation of the quantity $(C_{S} -\langle \partial_{-\beta} \hat{E}^{\ast}_S \rangle)=\text{Tr}[ (\frac{d \rho_{\infty}}{d-\beta}) \Delta \hat{E}^{\ast}_{S}]$, where $\Delta\hat{E}^{\ast}_{S}=\hat{E}^{\ast}_{S}-\langle \hat{E}^{\ast}_{S} \rangle$.  A correspondence between the $P$-representation of $\Delta \hat{E}^{\ast}_{S}$ and the score $\mathcal{L}_{\beta}$ is derived as
$\mathcal{L}_{\beta}(\Omega) Q_{\infty}(\Omega)=\int d\mu(\Omega) \mathcal{T}_{SI}(\Omega, \Omega^{\prime}) C_{\Delta E^{\ast}_{S}}^{(1)}(\Omega^{\prime})$, where $\mathcal{T}_{SI}=\int^{1}_{0} d\lambda \text{Tr}[\hat{\Pi}_{\Omega} \rho^{1-\lambda}_{\infty} \hat{\Pi}_{\Omega^\prime} \rho^{\lambda}_{\infty}]$ is the transformation function specific to SI.  An equilibrium state temperature uncertainty based on the SI is derived as
\begin{equation}
	\Delta \beta^{2} \langle (\delta\mathcal{H}_{tra}+\mathcal{H}_{bac} )^2\rangle_{g_{SI}}\ge 1,
\label{eq:TU-WDI}
\end{equation}
where $g_{SI}^{\Omega,\Omega^{\prime}} \equiv Q_{\infty}(\Omega) \mathcal{T}^{-1}_{SI} (\Omega,\Omega^{\prime}) Q_{\infty}(\Omega^{\prime})$ is the metric.   Generalizing it to non-equilibrium case, we can find that Eq. \eqref{eq:GTU} and Eq. \eqref{eq:TU-WDI} share the same expression, with only a difference in metric $g$. By recognizing both QFI and SI are Riemannian metrics of the Hilbert space \cite{Pires2016}, we can generalize Eq. \eqref{eq:GTU} to a universal case
\begin{equation}
	\Delta \beta^{2} \langle (\delta\mathcal{H}_{tra}+\mathcal{H}_{bac} )^2 \rangle_{g_{G}}\ge 1,
	\label{eq:Fluc-General}
\end{equation}
where $G$ can be arbitrary metric in Hilbert space.

In summary, we obtain a universal expression of temperature uncertainty relation for general non-equilibrium processes, applicable for any open quantum system, e.g., nanoscale systems or strong coupling systems. We find that it is the fluctuation of the trajectory heat plus backaction heat under measure $g$ that determines the temperature uncertainty.  Further analysis reveals that correlations between system and bath's process function (heat) and state function (energy) are the resources for decreasing the temperature uncertainty. Finally, by generalizing our conclusion to a more general case holding for any Riemannian metrics, the equilibrium state temperature uncertainty relation shown in Ref. \cite{Miller2018} is recovered.  More interestingly,  by relating QFI with the fluctuation of heat, our work connects the quantum thermodynamics and information theory together.  A study on this topic is our next project.

The authors thank Dr. Wei Wu and Professor Jun-Hong An for many fruitful discussions. The work was supported by the National Natural Science Foundation (Grant No. 12047501).

\bibliography{TU}

\end{document}